\documentclass[runningheads]{llncs}

\input{preamble.tex}

\begin{document}

\title{Efficient Graph Minors Theory and Parameterized Algorithms for (Planar) Disjoint Paths\thanks{This project has received funding from the European Research Council (ERC) under the European Union’s Horizon 2020 research and innovation programme (grant agreement no.~819416 and no.~715744). The second author also acknowledges the support of Swarnajayanti Fellowship grant DST/SJF/MSA-01/2017-18.  The third author acknowledges the support of ISF grant no.~1176/18.  The first and third authors also acknowledge the support of BSF grant no.~2018302.}}
\titlerunning{Efficient Graph Minors Theory}

\author{Daniel Lokshtanov\inst{1} \and
Saket Saurabh\inst{2} \and
Meirav Zehavi\inst{3}}
\authorrunning{D. Lokshtanov et al.}
%
\institute{University of California, Santa Barabara, USA.
\email{daniello@ucsb.edu}
\and
Indian Institute of Mathematical Sciences, Chennai, India, IRL 2000 ReLaX, and University of Bergen, Norway.
\email{saket@imsc.res.in}
 \and
Ben-Gurion University of the Negev, Beersheba, Israel.
\email{meiravze@.bgu.ac.il}}

\maketitle

\begin{abstract} 

In the {\sc Disjoint Paths} problem, the input consists of an $n$-vertex graph $G$ and a collection of $k$ vertex pairs, $\{(s_i,t_i)\}_{i=1}^k$, and the objective is to determine whether there exists a collection $\{P_i\}_{i=1}^k$ of $k$ pairwise vertex-disjoint paths in $G$ where the end-vertices of $P_i$ are $s_i$ and $t_i$. This problem was shown to admit an $f(k)n^3$-time algorithm by Robertson and Seymour {\em Graph Minors XIII, The Disjoint Paths Problem, JCTB}. In modern terminology, this means that {\sc Disjoint Paths} is fixed parameter tractable (FPT) with respect to $k$. Remarkably, the above algorithm for {\sc Disjoint Paths} is a cornerstone of the entire Graph Minors Theory, and conceptually vital to the  $g(k)n^3$-time algorithm for {\sc Minor Testing} (given two undirected graphs, $G$ and $H$ on $n$ and $k$ vertices, respectively, determine whether $G$ contains $H$ as a  minor).

In this semi-survey, we will first give an exposition of the Graph Minors Theory with emphasis on efficiency from the viewpoint of Parameterized Complexity.  Secondly, we will review the state of the art with respect to the {\sc Disjoint Paths} and {\sc Planar Disjoint Paths} problems. Lastly, we will discuss the main ideas behind a new algorithm that combines treewidth reduction and an algebraic approach to solve {\sc Planar Disjoint Paths} in time $2^{k^{\OO(1)}}n^{\OO(1)}$ (for undirected graphs).

\end{abstract}


\section{Background on Graph Minors Theory with Emphasis on Efficiency}\label{sec:intro}

Arguably, the origin of Parameterized Complexity is the {\em graph minors project of Robertson and Seymour}. Recollecting the birth of Parameterized Complexity, Downey \cite{DBLP:conf/birthday/Downey12} stated not only that ``a real inspiration was the theorem of Robertson and Seymour'', but also that in the early years of the field, ``many listeners thought that what we were doing was basically applying Robertson-Seymour.'' The concept of a minor originated already in the early 20th century. Formally, for any two graphs $G$ and $H$, we say that $H$ is a {\em minor} of $G$ if there exists a series of edge deletions, edge contractions and vertex deletions in $G$ that yields $H$. One of the most famous results in Graph Theory is Kuratowski's theorem~\cite{Kuratowski}, which states that a graph is planar if and only if it does not contain the graphs $K_{3,3}$ and $K_5$ as minors. That is, the class of planar graphs is characterized by a set of two forbidden minors. Robertson and Seymour set out to prove a vast generalization of Kuratowski's theorem, namely, Wagner's conjecture~\cite{Wagner}:
{\em Any infinite sequence of graphs contains two graphs such that one is a minor of the other (that is, the class of all graphs is well-quasi ordered by the minor relation)}.
Equivalently, Wagner's conjecture states that {\em any} minor-closed family of graphs can be characterized by a finite set of forbidden minors.

In perhaps one of the most amazing feats of modern mathematics, Robertson and Seymour managed to prove Wagner's conjecture. The endeavour of Robertson and Seymour to prove this conjecture spans a series of over 23 papers, published from 1983 to 2004. One of the main reasons why this theory has had such a great impact is the sheer number of novel algorithms and algorithmic techniques that were developed as a part of it. A few notable algorithmic highlights are their parameterized algorithms for {\sc Minor Testing} (given two graphs, $G$ on $n$ vertices and $H$ on $k$ vertices, decide whether $H$ is a minor of $G$), {\sc Disjoint Paths} (given a graph $G$ and a collection of $k$ terminal pairs, $\{(s_i,t_i)\}|_{i=1}^k$, decide whether $G$ has $k$ pairwise vertex-disjoint paths, $\{P_i\}|_{i=1}^k$, where for every $i\in\{1,2,\ldots,k\}$, the endpoints of $P_i$ are $s_i$ and $t_i$) and a constant-factor approximation parameterized algorithm to compute the treewidth of a given graph. Their project introduced key definitions and concepts such as those of an {\em excluded grid} and a {\em tree decomposition} (a decomposition of a graph into a tree-like structure), along with key structural results such as duality theorems (e.g., the characterization of treewidth in the terms of a family of connected graphs called a {\em bramble}). Additionally, their project presented new methods such as the so-called {\em irrelevant vertex technique}. 
Notably, {\em all} of these intermediate results have found applications and implications across a wide range of research~domains.

Unfortunately, the hidden constants in the results above, both in terms of time complexities and in the structural theorems themselves, are {\em really, really bad}. In fact, the immense parameter dependence of algorithms based on the graph minors project earned them their own name---``galactic algorithms''~\cite{lipton2013people}. As phrased by Johnson \cite{DBLP:journals/jal/Johnson87}, ``for any instance $G = (V,E)$ that one could fit into the known universe, one would easily prefer $|V|^{70}$ to even constant time, if that constant had to be one of Robertson and Seymour's''. In light of this, Downey \cite{DBLP:conf/birthday/Downey12} stated that ``in retrospect, it might have been a bit unfortunate to tie the FPT material to the Robertson-Seymour material when we spoke''. Indeed, keeping in mind that the primary objective of the paradigm of Parameterized Complexity is to cope with computational intractability, we are facing a blatant discrepancy:
\begin{quote}
{\bf While Parameterized Complexity does provide an extremely rich toolkit to design efficient parameterized algorithms, one of its foundations and still most powerful tools yields algorithms that are wildly impractical.}
\end{quote}

In 1989, Fellows \cite{richter1989graphs} noted that ``it is likely to be many years before the practical significance of Robertson-Seymour theorems is fully understood''. Nevertheless, for some of the algorithms, substantial advances have been made. In particular, Grohe et al.~\cite{GroheKR13} gave an algorithm for computing {\em Robertson and Seymour's structural decomposition} (stating that all graphs excluding some fixed graph as a minor have a tree decomposition with bags that are almost embeddable in a fixed surface), 
that runs in time $f(k)n^2$ (for some function $f$ of $k$), improving over the $f(k)n^3$-time algorithm of Robertson and Seymour.
Prior to this result, Kawarabayashi and Wollan~\cite{KawarabayashiW10} gave a simplified proof of correctness of the graph minors algorithm. This proof yields a parametrized algorithm for {\sc Minor Testing} with the best currently known parameter dependence. Here, $f(k)$ is a ``tower of powers of $2$ of height at most $5$, with $k^{1000}$ on top''~\cite{Wol15pers}. From the work of Kawarabayashi et al.~\cite{DBLP:journals/jct/KawarabayashiKR12} on {\sc Disjoint Paths}, we also know that {\sc Minor Testing} is solvable in time $f(k)n^2$.
 Chuzhoy~\cite{Chuzhoy14}, building upon the seminal work of Chekuri and Chuzhoy~\cite{ChekuriC14}, gave an improved algorithm for a weaker variant of Robertson and Seymour's structural decomposition. However, here the improvement is in the quality of the output decomposition, not in the time it takes to compute it.
 
Both the algorithm for {\sc Minor Testing} and the structural theorem of Robertson and Seymour were discovered as consequences of the entire graph minors theory that they built. In retrospect, however, the ``converse'' also holds true: almost the entire graph minors theory would have had to be built in order to achieve either one of these goals, that is, the algorithm for {\sc Minor Testing} as well as the structural theorem of Robertson and Seymour! In plain words, the design of an efficient algorithm for {\sc Minor Testing} as a goal {\em necessitates} to devise efficient versions of large parts of the whole graph minors theory. A problem that is tightly linked to {\sc Minor Testing}, yet seemingly more difficult than it, is {\sc Topological Minor Testing}: given two graphs, $G$ on $n$ vertices and $H$ on $k$ vertices, decide whether $H$ is a {\em topological minor} of $G$. That is, the objective is to determine whether there exists a series of operations that delete an edge, {\em dissolve} an edge (i.e. delete a degree-2 vertex and make its two neighbors adjacent) and delete a vertex in $G$ that yields $H$. We remark that Kuratowski's theorem~\cite{Kuratowski} was, in fact, originally phrased in the terms of topological minors rather than minors. Unlike {\sc Minor Testing} and {\sc Disjoint Paths}, the question of the parameterized complexity of {\sc Topological Minor Testing} was not resolved by Robertson and Seymour, and was first stated explicitly as an open problem by Downey and Fellows in 1992~\cite{DBLP:conf/coco/DowneyF92}. Since then, this question was restated as an open problem many times, until it was positively resolved by Grohe et al.~\cite{DBLP:conf/stoc/GroheKMW11}, who designed an $f(k)n^3$-time (galactic) algorithm (see also \cite{tmcNew}).

\subsection*{Some Central Applications}

Graph minors in general, and {\sc Minor Testing} in particular, have enjoyed numerous applications over the past 30 years. These applications span a wide range of areas, including (but not limited to) Approximation Algorithms, Exact Exponential and Polynomial-Time Algorithms, Parameterized Complexity, Logic, Computational Geometry and Property Testing. It is highly conceivable that algorithmically (and structurally) efficient Graph Minors Theory, being a core engine behind all of these applications, will have great impact on all of these areas simultaneously. For the sake of illustration, we briefly discuss three important discoveries (with emphasis on Parameterized Complexity) that build upon graph minors. 

\medskip
\noindent{\bf Classification.} By Robertson and Seymour's theorem, every minor-closed family of graphs can be characterized by a finite set of forbidden minors. In particular, for any minor-closed family of graphs $\cal G$, the family of graphs obtained from the graphs in $\cal G$ by adding at most $k$ vertices is also minor-closed. As {\sc Minor Testing} is solvable in time $f(k)n^2$~\cite{DBLP:journals/jct/KawarabayashiKR12}, this observation immediately shows that a vast range of parametrized problems, such as {\sc Feedback Vertex Set}, {\sc Planar Vertex Deletion} and {\sc Graph Genus}, are {\em non-uniformly \FPT}: for every integer $k$, the set of forbidden minors might be different. Furthermore, the result is non-constructive as long as we do not know how to compute the finite set of forbidden minors. Nevertheless, this result provides a very useful classification tool (for whether a problem is \FPT\ or not), known since the early days of Parameterized Complexity. 

\medskip
\noindent{\bf Bidimensionality.} The Graph Minors Theory laid the foundation for studying how computational problems that are hard on general graphs behave when restricted to minor-free graphs. The theory of {\em bidimensionality}~\cite{demaine2005subexponential} builds upon this knowledge, particularly on the relationship between grids and treewidth. While most \NP-hard graph problems remain \NP-hard even on planar graphs~\cite{GJ79},  many problems that are fixed-parameter intractable on general graphs are \FPT\ on planar graphs, and even on graph classes excluding a fixed $H$ as a minor. Bidimensionality simultaneously yields linear-time parameterized algorithms with subexponential parameter dependence~\cite{demaine2005subexponential}, polynomial-time approximation schemes \cite{DemaineH05,FominLRS11}, and linear kernels~\cite{FominLST10} for many problems on minor-free graphs, with applications even in computational geometry~\cite{DemaineFHT05Map,FominLS12}. Nevertheless, there are fundamental graph problems on planar and minor-free graphs for which bidimensionality seems insufficient. Examples of such problems include the {\sc Longest Path} problem on directed graphs, {\sc Steiner Tree}, {\sc Odd Cycle Transversal}, and many others~\cite{FellowsFLRSV12,LokshtanovSW12,PilipczukPSL14}. More information can be found in two other chapters in this volume (one by D.~Marx, and the other by Ma.~Pilipczuk).

\medskip
\noindent{\bf Irrelevant Vertices.}  The irrelevant vertex technique originated from Robertson and Seymour's algorithm for the {\sc Disjoint Paths} problem~\cite{DBLP:journals/jct/RobertsonS95b}. Since then, this technique has found several other applications~\cite{AdlerKKLST11,DBLP:conf/focs/CyganMPP13,DBLP:journals/tcs/GolovachHP13,DBLP:conf/stoc/GroheKMW11,DBLP:journals/algorithmica/Marx10}. Roughly speaking, as long as the treewidth of the graph is large, the technique is applied by repeatedly finding an {\em irrelevant vertex}---a vertex whose deletion does not change the answer to the problem. For an illustrative application of this technique in textbook level of detail, we refer to Chapter 7.8 in \cite{DBLP:books/sp/CyganFKLMPPS15}.

\section{(Planar) Disjoint Paths: State of the Art}\label{sec:state}

Conceptually vital to the algorithm for {\sc Minor Testing} of Robertson and Seymour, and the source of the irrelevant vertex technique, is their $f(k)n^3$-time algorithm for the {\sc Disjoint Paths} problem~\cite{DBLP:journals/jct/RobertsonS95b}. The current state-of-the-art is the algorithm developed by Kawarabayashi et al.~\cite{DBLP:journals/jct/KawarabayashiKR12} in 2012, which runs in time $f(k)n^2$. Just like the case of {\sc Minor Testing}, the parameter dependence of this algorithm on $k$ renders it a ``galactic algorithm''. The {\sc Disjoint Paths} problem is important on its own right due to its applications in the contexts of transportation networks, VLSI layout and virtual circuit routing~\cite{frank1990packing,schrijver2003combinatorial,DBLP:journals/tit/OgierRS93,DBLP:journals/winet/SrinivasM05}. It was shown to be \NP-complete by Karp in 1975~\cite{Karp:1975:CCC:3160092.3160096}, being one of Karp's original \NP-complete problems. In fact, it remains NP-complete even if the input graph is restricted to be a grid \cite{kramer1984complexity}. 

\subsection{Minor Testing and Disjoint Paths on Planar Graphs}
We first remark that the {\sc Minor Testing}, {\sc Disjoint Paths} and {\sc Topological Minor Testing} problems are well known to be \NP-hard also when restricted to the class of planar graphs~\cite{lynch1975equivalence}. Moreover, it is easy to see that if {\sc (Topological) Minor Testing} is solvable in time $2^{k^{\OO(1)}}n^c$ (for some $c>0$) on $H$-minor free graphs, then {\sc Disjoint Paths} is also solvable in time $2^{k^{\OO(1)}}n^c$ (for the same $c$) on this class of graphs. To see this, consider an instance $(G,\{(s_i,t_i)\}|_{i=1}^k,k)$ of {\sc Disjoint Paths} on $H$-minor free graphs. Briefly, the idea is to attach, to each terminal $s_i$ or $t_i$, a ``large enough'' clique (on $\OO(|V(H)|k)$ vertices) of unique size, and define the graph to be sought as a minor as an appropriate combination of these cliques. Unfortunately, this reduction idea is clearly tailored specifically to $H$-minor free graphs---in particular, it is inapplicable to planar graphs and general graphs.

Apart from being important problems on their own right, the design of algorithms for {\sc Minor Testing}, {\sc Disjoint Paths} and {\sc Topological Minor Testing} on planar graphs also serves as a critical building block for the design of algorithms for these problems on general graphs in view of the way Robertson and Seymour's Graph Minors Theory is structured. Without delving into technical details, we note that all known algorithms for {\sc (Topological) Minor Testing} and {\sc Disjoint Paths} (on general graphs) are based on the distinction between the case where the input graph $G$ contains a large clique as a minor, and the case where it does not. Already at this stage, we see that the resolution of {\sc (Topological) Minor Testing} and {\sc Disjoint Paths} on $H$-minor-free graphs is a building block towards the resolution of these problems on general graphs. Moreover, when the input graph does not contain a large clique as a minor, known algorithms distinguish between the case where the treewidth of $G$ is small, and the case where the treewidth of $G$ is large. In the latter case, $G$ contains a so-called {\em flat wall} that further motivates, or even necessitates, the study of these problems on planar and ``almost planar'' graph classes.

With respect to known algorithms, for the {\sc Minor Testing} problem on planar graphs, the design of a $2^{k^{\OO(1)}}n$-time algorithm is folklore: if the input graph $G$ has treewidth larger than some function linear in $k$, then it necessarily contains the sought graph as a minor (this is a property holds only for planar graphs!), and otherwise it is possible to solve the problem in time $2^{k^{\OO(1)}}n$ via dynamic programming over tree decompositions (e.g., using \cite{DBLP:journals/tcs/AdlerDFST11}).
We also remark that, for {\sc Minor Testing} on planar graphs, Adler et al.~\cite{DBLP:journals/algorithmica/AdlerDFST12} developed an algorithm that runs in time $\OO(2^{\OO(k)}n + n^2\log n)$.

For the {\sc Disjoint Paths} problem on planar graphs~\cite{DBLP:journals/dam/Reed95,DBLP:conf/gst/ReedRSS91}, and even on graphs of bounded genus~\cite{DBLP:journals/dam/Reed95,DBLP:conf/soda/DvorakKT09,DBLP:conf/soda/KobayashiK09}, there already exist algorithms with running times whose dependency on $n$ in linear, but whose dependency on $k$ is prohibitive. Additionally, for the {\sc Disjoint Paths} problem on planar graphs, Adler et al.~\cite{DBLP:journals/jct/AdlerKKLST17} developed a $2^{2^{\OO(k)}}n^2$-time algorithm; in particular, towards that end, they presented a so-called {\em unique linkage theorem} that states that, in every instance of {\sc Disjoint Paths} on planar graphs where the treewidth is larger than $2^{ck}$ (for some $c>0$), there exists an irrelevant vertex and it is computable in linear time. Such a relation with single exponential dependency of the treewidth bound on $k$ also holds for graphs of bounded genus~\cite{mazoit2013single}. However, for the more general $H$-minor-free graphs, the dependency becomes a tower of exponents~\cite{DBLP:journals/jct/GeelenHR18,KawarabayashiW11} (prior to these works---that is, from the project of Robertson and Seymour---not even the computability of the bound was not known).

The {\sc Planar Disjoint Paths} problem is intensively studied also from the perspective of approximation algorithms, with a burst of activity in recent years \cite{ChuzhoyK15,ChuzhoyKL16,ChuzhoyKN17,ChuzhoyKN18,DBLP:conf/icalp/ChuzhoyKN18}. Highlights of this work include an approximation algorithm with approximation factor $\cO(n^{9/19} \log ^{\cO(1)}n)$~\cite{ChuzhoyKL16} and, under reasonable complexity-theoretic assumptions, the proof of hardness of approximating the problem within a factor of $2^{\Omega{( \frac{1}{(\log \log n)^2} ) } }$~{\cite{ChuzhoyKN18}.
For the {\sc Directed Disjoint Paths} problem on planar graphs, Schrijver~\cite{DBLP:journals/siamcomp/Schrijver94} gave an algorithm with running time $n^{\cO(k)}$, in contrast to the NP-hardness for $k=2$ on general directed graphs. Almost 20 years later, Cygan et al.~\cite{DBLP:conf/focs/CyganMPP13} improved over the algorithm of Schrijver and showed that {\sc Directed Disjoint Paths} on planar graphs is FPT by giving an algorithm with running time $2^{2^{\cO(k^2)}} n^{\cO(1)}$.

\subsection{General Structure of Algorithms for (Planar) Disjoint Paths}\label{sec:genStruct}

All known algorithms for both {\sc Disjoint Paths} and {\sc Planar Disjoint Paths} have the same high level structure. In particular, given a graph $G$ we distinguish between the cases of $G$ having ``small'' or ``large'' treewidth. In case  the treewidth is large, we distinguish between two further cases: either $G$ contains a ``large'' clique minor or it does not. This results in the following case distinctions. 

\begin{enumerate}
\item {\bf Treewidth is small.} Let the treewidth of $G$ be $w$. Then, we use the known dynamic programming algorithm with running time 
$2^{\cO(w \log w)}n^{\cO(1)}$~\cite{scheffler1994practical} to solve the problem.  
It is important  to note that, assuming the Exponential Time Hypothesis (ETH), there is neither an algorithm for {\sc Disjoint Paths}  running in time $2^{o(w \log w)}n^{\cO(1)}$ \cite{DBLP:journals/siamcomp/LokshtanovMS18}, 
nor an algorithm for {\sc Planar Disjoint Paths} running in time $2^{o(w)}n^{\cO(1)}$~\cite{DBLP:journals/tcs/BasteS15}. 

\item {\bf Treewidth is large and $G$ has a large clique minor.} In this case, we use the good routing property of the clique to find an irrelevant vertex and delete it without changing the answer to the problem. Since this case does not arise for graphs embedded on a surface or for planar graphs, we do not discuss it in more detail. 

\item {\bf  Treewidth is large and $G$  has no large clique minor.} Using a fundamental structure theorem for minors called the flat wall theorem,  we can conclude that $G$ contains a large planar piece and a vertex $v$ that is sufficiently insulated in the middle of it. Applying the unique linkage theorem~\cite{DBLP:journals/jct/RobertsonS12}  to this vertex, we conclude that it is irrelevant and remove it. For planar graphs, one can use the unique linkage theorem of Adler et al.~\cite{DBLP:journals/jct/AdlerKKLST17}:
\begin{quote}
Any instance of {\sc Disjoint Paths} consisting of a planar graph with treewidth at least $82 k^{3/2}2^k$ and $k$ terminal pairs contains a vertex $v$ such that every solution to {\sc Disjoint Paths}   can be replaced by an equivalent one whose paths avoid $v$.

\end{quote}
This result says that if the treewidth of the input planar graph is (roughly) $\Omega(2^k)$, then we can find an irrelevant vertex and remove it. A natural question is whether we can guarantee an irrelevant vertex even if the treewidth is $\Omega({\sf poly}(k))$. Adler and Krause~\cite{DBLP:journals/corr/abs-1011-2136} exhibited a planar graph $G$ with $k+1$ terminal pairs such that $G$ contains a $(2^k + 1) \times (2^k + 1)$ grid as a subgraph,  {\sc Disjoint Paths} on this input has a unique solution, and  the solution uses all vertices of $G$; in particular, {\em no vertex of $G$ is irrelevant}. This implies that the irrelevant vertex technique can only guarantee a treewidth of $\Omega(2^k)$, even if the input graph is planar. 
\end{enumerate}

\noindent Combining items (1) and (3), observe that the known methodology for {\sc Disjoint Paths}  can only guarantee an algorithm with running time $2^{2^{\cO(k)}}n^2$ even when restricted to planar graphs. Thus, a $2^{k^{\OO(1)}}n^{\cO(1)}$-time algorithm for {\sc Planar Disjoint Paths} appears to require entirely new ideas. As this obstacle was known to Adler et al.~\cite{AdlerOpen13}, it is likely to be the main motivation for Adler to pose the existence of a $2^{k^{\OO(1)}}n^{\cO(1)}$ time algorithm for {\sc Planar Disjoint Paths} as an open problem. 


\section{Recent Development: Combination of Treewidth Reduction and an Algebraic Approach}\label{sec:combination}

Recent joint work of the authors of this paper with Misra and Pilipczuk~\cite{newPDP} led to the development of an FPT algorithm for {\sc Planar Disjoint Paths} whose parameter dependency on $k$ is bounded by $2^{\OO(k^2)}$. Specifically, we proved the following theorem.

\begin{thm}[\cite{newPDP}]\label{thm:main}
The {\sc Planar Disjoint Paths} problem is solvable in time $2^{\OO(k^2)}n^{\OO(1)}$.
\end{thm}

Our algorithm is based on a novel combination of two techniques that do not seem to give the desired outcome when used on their own. The first ingredient is the treewidth reduction theorem of Adler et al.~\cite{DBLP:journals/jct/AdlerKKLST17} that proves that given an instance of {\sc Planar Disjoint Paths}, the treewidth can be brought down to $2^{\cO(k)}$ (see item (3) in Section \ref{sec:genStruct}).  This by itself is sufficient for an FPT algorithm (this is what Adler et al.~\cite{DBLP:journals/jct/AdlerKKLST17} do), but as explained above, it seems hopeless that it will bring a $2^{k^{\cO(1)}}n^{\cO(1)}$-time algorithm. 

We circumvent the obstacle by using  an algorithm for a more difficult problem with a worse running time, namely,  Schrijver's $n^{\cO(k )}$-time algorithm for {\sc Disjoint Paths} on directed planar graphs~\cite{DBLP:journals/siamcomp/Schrijver94}. Schrijver\rq{}s algorithm has two steps: a ``guessing'' step where one (essentially) guesses the homology class of the solution paths, and then a surprising homology-based algorithm that, given a homology class, finds a solution in that class (if one exists) in polynomial time. Our key insight is that for {\sc Planar Disjoint Paths}, if the instance that we are considering has been reduced according to the procedure of Adler et al.~\cite{DBLP:journals/jct/AdlerKKLST17}, then we only need to iterate over $2^{\cO(k^2)}$ homology classes in order to find the homology class of a solution, if one exists. The proof of this key insight is highly non-trivial, and builds on a cornerstone ingredient of the FPT algorithm of Cygan et al.~\cite{DBLP:conf/focs/CyganMPP13} for {\sc Directed Disjoint Paths} on planar graphs. To the best of our knowledge, this is the first algorithm that finds the exact solution to a problem that exploits that the treewidth of the input graph is small in a way that is different from doing dynamic programming.

In what follows, we begin with an explanation of the statement of the main technical result of Schrijver~\cite{DBLP:journals/siamcomp/Schrijver94}. Then, we will introduce a special Steiner tree that is one of the key components in the design of our algorithm, in order to explain one of the arguments where treewidth reduction is critical.

\subsection{Schrijver's Main Technical Result}

The starting point of our algorithm is Schrijver's view \cite{DBLP:journals/siamcomp/Schrijver94} of a collection of ``non-crossing'' (but possibly not vertex- or even edge-disjoint) sets of walks as flows. 
To work with flows (defined immediately), we deal with directed graphs. (In this context, undirected graphs are treated  as directed graphs by replacing each edge by two parallel arcs of opposite directions.) Specifically, we denote an instance of {\sc Directed Planar Disjoint Paths} as a tuple $(D,S,T,g,k)$ where $D$ is a directed plane graph, $S,T\subseteq V(D)$, $k=|S|$ and $g: S\rightarrow T$ is bijective. Then, a {\em solution} is a set ${\cal P}$ of pairwise vertex-disjoint directed paths in $D$ containing, for each vertex $s\in S$, a path directed from $s$ to $g(s)$.
In the language of flows, each arc of $D$ is assigned a word with letters in $T\cup T^{-1}$ (that is, we treat the set of vertices $T$ also as an alphabet), where $T^{-1}=\{t^{-1}: t\in T\}$. A word is {\em reduced} if, for all $t\in T$, the letters $t$ and $t^{-1}$ do not appear consecutively. Then, a {\em flow} is an assignment of reduced words to arcs that satisfies two constraints. First, when we concatenate the words assigned to the arcs incident to a vertex $v\notin S\cup T$ in clockwise order, where words assigned to ingoing arcs are reversed and their letters negated, the result (when reduced) is the empty word (see Fig.~\ref{fig:flowStitch}). This is an algebraic interpretation of the standard flow-conservation constraint. Second, when we do the same operation with respect to a vertex $v\in S\cup T$, then when the vertex is in $S$, the result is $g(s)$ (rather than the empty word), and when it is in $T$, the result is $t$. There is a natural association of flows to solutions: for every $t\in T$, assign the letter $t$ to all arcs used by the path from $g^{-1}(t)$ to $t$.

\begin{figure}
    \begin{center}
        \includegraphics[width=0.45\textwidth]{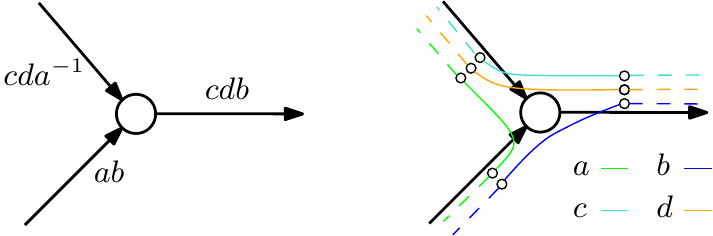}
        \caption{Flow at a vertex and its reduction.}
        \label{fig:flowStitch}
    \end{center}
\end{figure}

Roughly speaking, Schrijver proved that if a flow $\phi$ is given along with the instance $(D,S,T,g,k)$, then in {\em polynomial time} we can either find a solution or determine that there is no solution ``similar to $\phi$''. Specifically, two flows are {\em homologous} (which is the notion of similarity) if one can be obtained from the other by a {\em set} of ``face operations'' defined as follows.

\begin{defe}\label{def:homologyOverview}
Let $D$ be a directed plane graph with outer face $f$, and denote the set of faces of $D$ by $\cal F$. Two flows $\phi$ and $\psi$ are {\em homologous} if there exists a function $h: {\cal F}\rightarrow (T\cup T^{-1})^*$ such that {\em (i)} $h(f)=1$, and {\em (ii)} for every arc $e\in A(D)$, $h(f_1)^{-1}\cdot \phi(e)\cdot h(f_2)=\psi(e)$ where $f_1$ and $f_2$ are the faces at the left-hand side and the right-hand side of $e$, respectively.
\end{defe} 

Then, a slight modification of Schrijver's theorem~\cite{DBLP:journals/siamcomp/Schrijver94} yields the following corollary.

\begin{coro}\label{prop:schOverview}
There is a polynomial-time algorithm that, given an instance $(D,$ $S,T,g,k)$ of {\sc Directed Planar Disjoint Paths}, a flow $\phi$, and a subset $X\subseteq A(D)$, either finds a solution of $(D-X,S,T,g,k)$ or decides that there is no solution of it such that the ``flow associated with it'' and $\phi$ are homologous in $D$.
\end{coro}

\begin{figure}
    \begin{center}
        \includegraphics[scale=0.375]{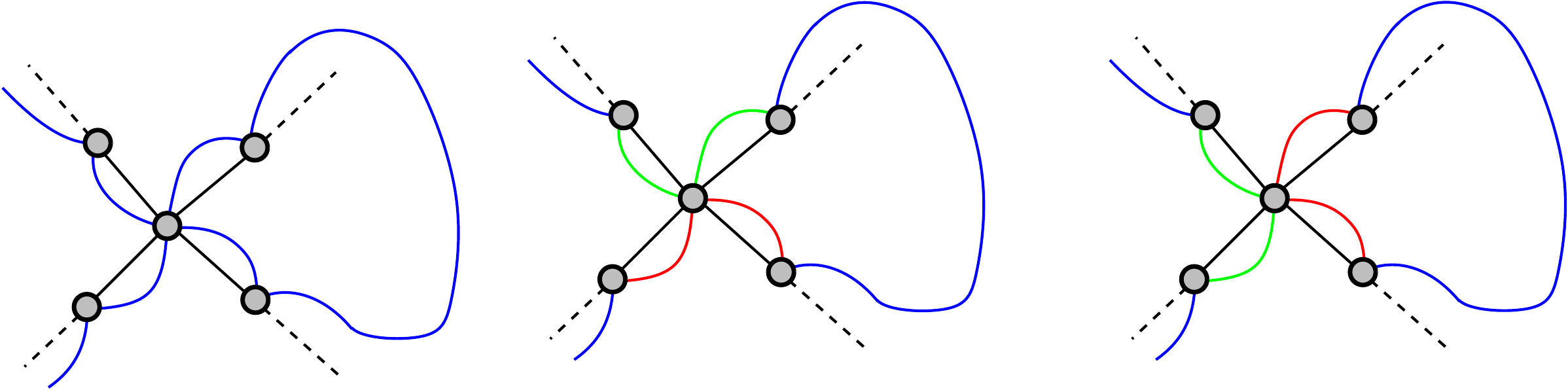}
        \caption{Two different ways of extracting a walk from a flow.}
        \label{fig13}
    \end{center}
\end{figure} 

\medskip
\noindent{\bf Discrete Homotopy and Our Objective.} The language of flows brings several technicalities such as having different sets of non-crossing walks corresponding to the same flow (see Fig.~\ref{fig13}). Instead, we may define a notion of {\em discrete homotopy}, which is an equivalence relation that consists of three face operations. Then, we deal only with collections of non-crossing edge-disjoint walks, called {\em weak linkages}.  Roughly speaking, two weak linkages are {\em discretely homotopic} if one can be obtained from the other by using ``face operations'' that push/stretch its walks across faces and keep them non-crossing and edge-disjoint (see Fig.~\ref{fig0203}). We note that the order in which face operations are applied is important in discrete homotopy (unlike homology)---e.g., we cannot stretch a walk across a face if no walk passes its boundary, but we can execute operations that will move a walk to that face, and then stretch it. We can translate Corollary \ref{prop:schOverview} to discrete homotopy (and undirected graphs) to derive the following result.

\begin{lem}\label{lem:discreteHomotopyOverview}
There is a polynomial-time algorithm that, given an instance $(G,S,$ $T,g,k)$ of {\sc Planar Disjoint Paths}, a weak linkage $\cal W$ in $G$ and a subset $X\subseteq E(G)$, either finds a solution of $(G-X,S,T,g,k)$ or decides that no solution of it is discretely homotopic to $\cal W$ in $G$.
\end{lem}

\begin{figure}
    \begin{center}
        \includegraphics[width=0.475\textwidth]{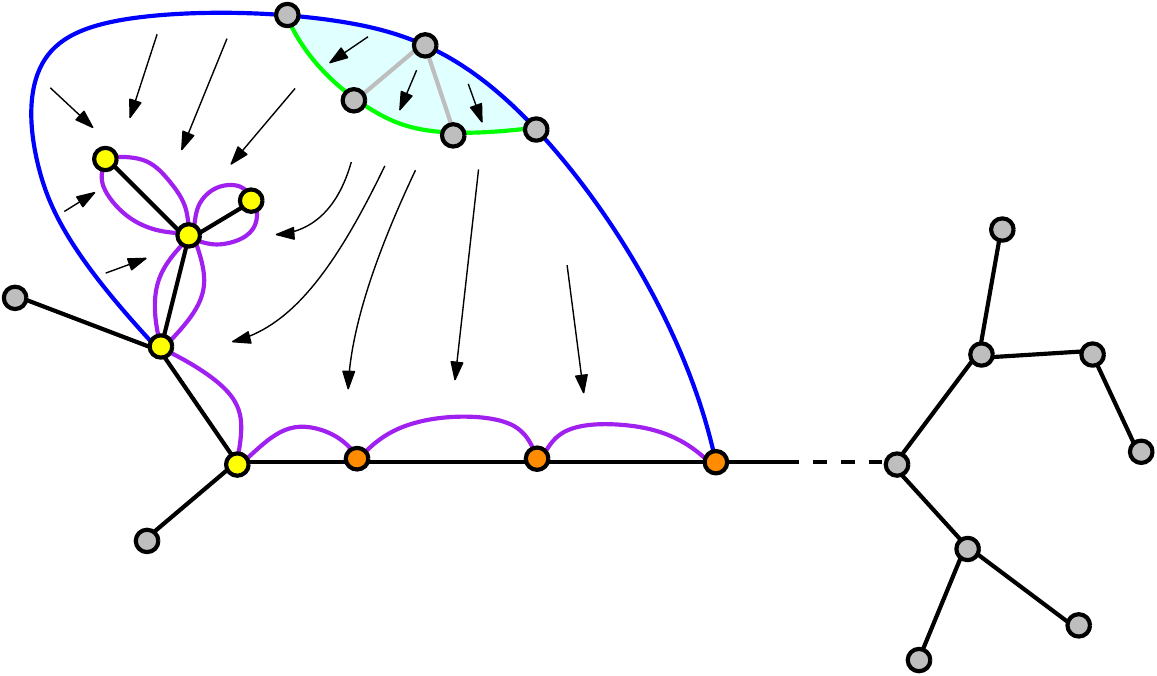}
        \caption{Moving a weak linkage (having one walk) with ``face operations''.}
        \label{fig0203}
    \end{center}
\end{figure}

In light of this result, our objective is reduced to the following task.

\begin{quote}
\framebox{
\begin{minipage}{0.85\textwidth}
Compute a collection of weak linkages such that if there exists a solution, then there also exists a solution ({\em possibly a different one!}) that is discretely homotopic to one of the weak linkages in our collection. To prove Theorem \ref{thm:main}, the size of the collection should be upper bounded by $2^{\OO(k^2)}$.
\end{minipage}
}
\end{quote}

This task turns out to be very challenging, and our current manuscript spans roughly 80 pages to achieve it. In the next section, we will describe one of the main ingredients, which also allows us to hint at the necessity of treewidth reduction at preprocessing.

\subsection{Key Player: Steiner Tree}

\noindent A key to the proof of our theorem is a very careful construction (done in three steps) of a so-called {\em Backbone Steiner tree} $R$. We use the term Steiner tree to refer to any tree in the {\em radial completion} of $G$ (the graph obtained by placing a vertex on each face and making it adjacent to all vertices incident to the face) whose set of leaves is precisely $S\cup T$.  Having the aforementioned Backbone Steiner tree $R$ at hand, we have a more focused goal: we will zoom into weak linkages that are ``pushed onto $R$'', and we will only generate such weak linkages to construct our collection. Informally, a weak linkage is {\em pushed onto $R$} if all of the edges used by all of its walks are {\em parallel to} edges of $R$. We do not demand that the edges belong to $R$ itself, because then the aforementioned goal cannot be achieved---some edges will have to be used several times, which prevents satisfying the edge disjointness requirement. Instead, we make $\Theta(n)$ parallel copies of each edge in the radial completion (the precise number  arises from considerations in the ``pushing process''), and then impose the weaker demand of being parallel. Now, our goal is to show the following statement: If there exists a solution, then there also exists one that can be pushed onto $R$ by applying face operations (in discrete homotopy) so that it becomes {\em identical} to one of the weak linkages in our collection (see Fig.~\ref{fig0203}).

At this point, one remark is in place. Our Steiner tree $R$ is a subtree of the radial completion of $G$ rather than $G$ itself. Thus, if there exists a solution discretely homotopic to one of the weak linkages that we generate, it might not be a solution in $G$. We easily circumvent this issue by letting the set $X$ in Lemma \ref{lem:discreteHomotopyOverview} contain all ``fake'' edges.

\medskip
\noindent{\bf Example of the Necessity of Treewidth Reduction.} To be able to generate a collection of only $2^{k^{\OO(1)}}$ weak linkages so that the aforementioned statement can be proven, we carefully construct our Backbone Steiner tree $R$ in three steps so that it will satisfy several critical properties. (The third step is the most technical one, and will not be presented here). The first step is merely an initialization step, where we consider an arbitrary Steiner tree as $R$. Then, in the second step we modify $R$ as follows. For each ``long'' maximal degree-2 path $P$  of $R$ with endpoints $u$ and $v$, we will compute two minimum-size vertex sets, $S_u$ and $S_v$, such that $S_u$ separates (i.e., intersects all paths between) the following two subgraphs in the radial completion of $G$: {\em (i)} the subtree of $R$ that contains $u$ after the removal of a vertex $u_1$ of $P$ that is ``very close'' to $u$, and {\em (ii)} the subtree of $R$ that contains $v$ after the removal of a vertex $u_2$ that is ``close'' to $u$. The condition satisfied by $S_v$ is symmetric (see Fig.~\ref{fig05}). Here, ``very close'' refers to distance $2^{c_1k}$ and ``close'' refers to distance $2^{c_2k}$ for some constants $c_1<c_2$. (The selection of $u'$ not to be $u$ itself is of use in the third modification of $R$.)

\begin{figure}
    \begin{center}
        \includegraphics[width=0.875\textwidth]{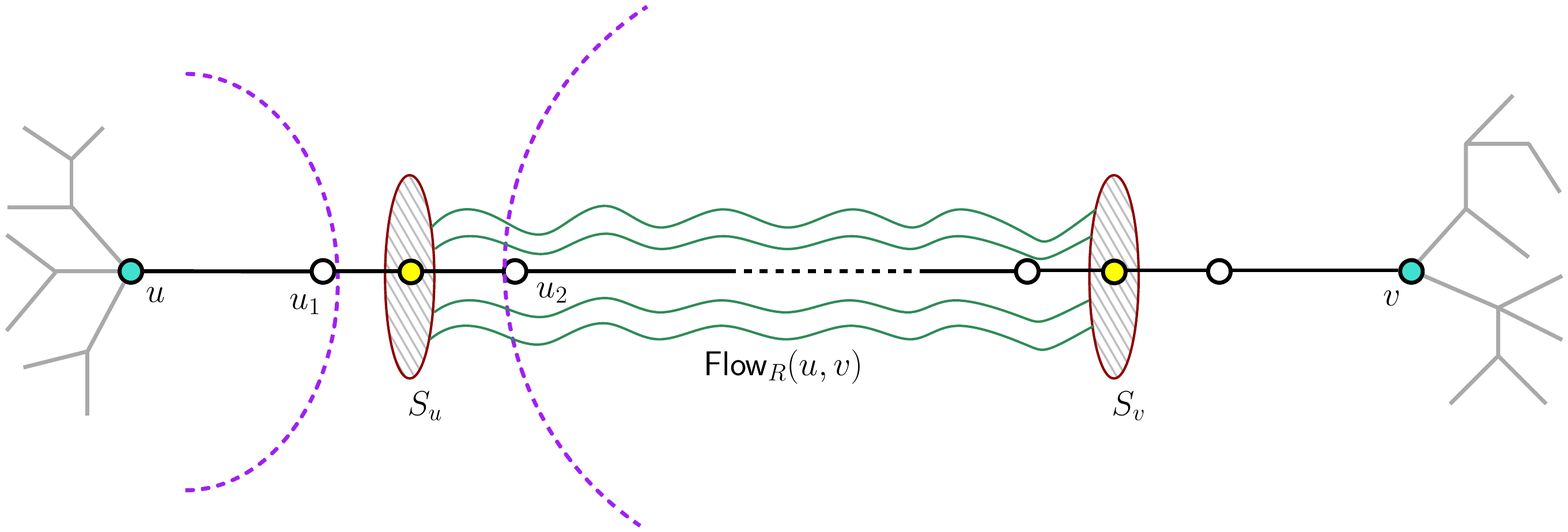}
        \caption{Separators and flows for a long maximal degree-2 path $P$ in $R$.}
        \label{fig05}
    \end{center}
\end{figure}

To utilize these separators, we need their sizes to be upper bounded by $2^{\OO(k)}$. For our initial Steiner tree $R$, such small separators may not exist. However, the modification we will present now can be shown to guarantee their existence. Specifically, we will ensure that $R$ does not have any {\em detour}, which roughly means that each of its maximal degree-2 paths is a shortest path connecting the two subtrees obtained once it is removed. More formally, we define a detour as follows (see Fig.~\ref{fig:undetour}).

\begin{figure}
    \begin{center}
        \includegraphics[scale=0.475]{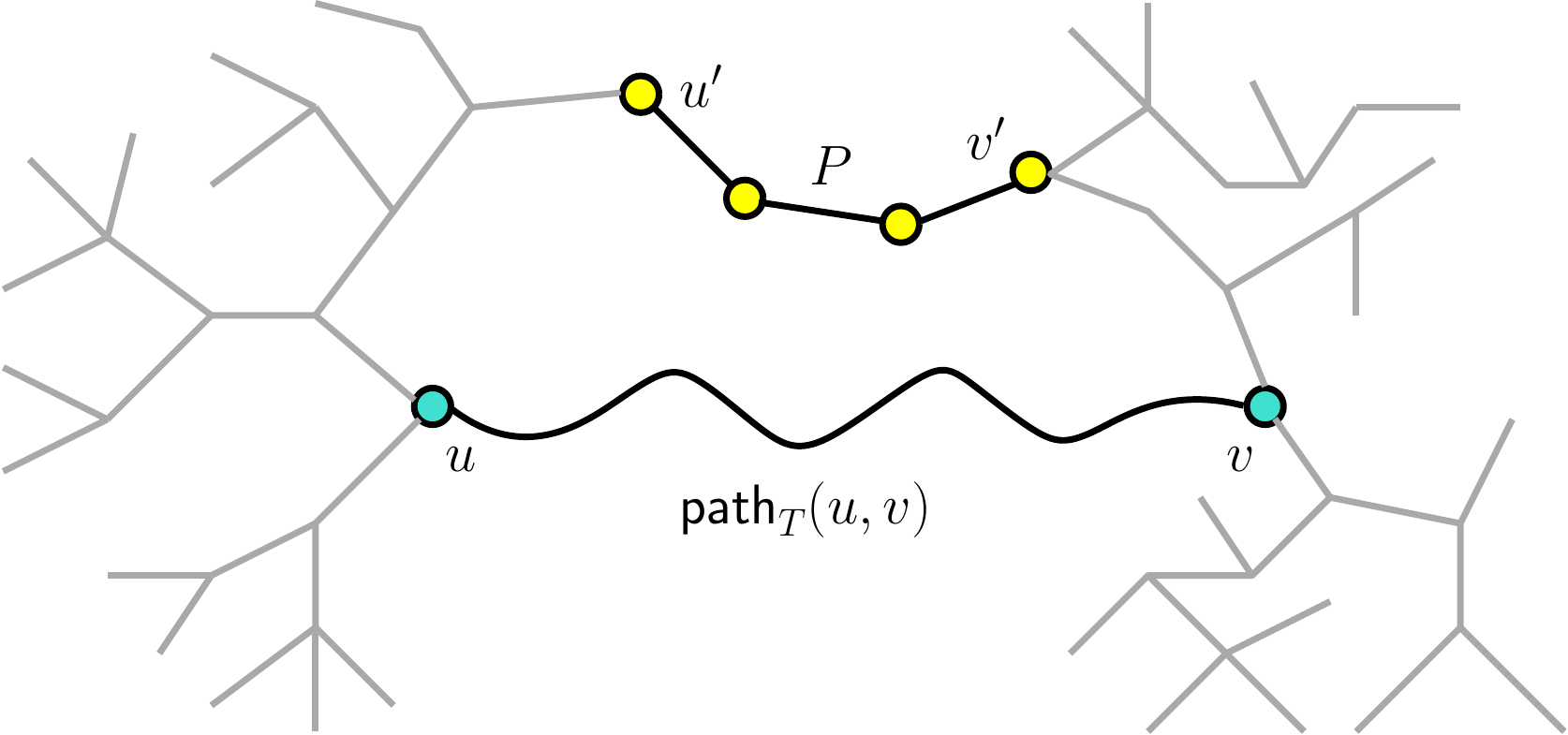}
        \caption{A detour in a Steiner tree $T$.}
        \label{fig:undetour}
    \end{center}
\end{figure}

\begin{defe}\label{def:detourOverview}
A {\em detour} in $R$ is a pair of vertices $u,v\in V_{\geq 3}(R)\cup V_{=1}(R)$ that are endpoints of a maximal degree-2 path $L$ of $R$, and a path $P$ in the radial completion of $G$, such that {\em (i)} $P$ is shorter than $L$, {\em (ii)} one endpoint of $P$ belongs to the component of $R-V(L)\setminus\{u,v\}$ containing $u$, and {\em (iii)} one endpoint of $P$ belongs to the component of $R-V(L)\setminus\{u,v\}$ containing $v$.
\end{defe}

By repeatedly ``short-cutting'' $R$, a process that terminates in a linear number of steps, we obtain a new Steiner tree $R$ with no detour. Now, if the separator $S_u$ is large, then there is a large number of vertex-disjoint paths that connect the two subtrees separated by $S_u$, and all of these paths are ``long'', namely, of length at least $2^{c_2k}-2^{c_1k}$. Based on a result by Bodlaender et al.~\cite{DBLP:journals/jacm/BodlaenderFLPST16} (whose application requires to work in the radial completion of $G$ rather than $G$ itself), we can show that the existence of these paths implies that the treewidth of $G$ is large. Thus, if the treewidth of $G$ were small, all of our separators would have also been small. Fortunately, to guarantee this, we just need to invoke the known treewidth reduction for {\sc Planar Disjoint Paths} in a preprocessing step: 

\begin{prop}[\cite{DBLP:journals/jct/AdlerKKLST17}]\label{prop:twReductionOverview}
There is a $2^{\OO(k)}n^2$-time algorithm that, given an instance $(G,ST,g,k)$ of {\sc Planar Disjoint Paths}, outputs an equivalent instance $(G',$ $S,T,g,k)$ of {\sc Planar Disjoint Paths} where $G'$ is a subgraph of $G$ whose treewidth is upper bounded by $2^{ck}$ for some constant $c$.
\end{prop}

Intuitively, having separators of size $2^{\OO(k)}$ is useful with respect to the proof of sufficiency of generating a collection of only $2^{k^{\OO(1)}}$ weak linkages as follows. When we consider a solution $\cal P$ (if one exists), then from the fact that the paths in $\cal P$ are vertex disjoint, we immediately have the property that the paths in $\cal P$ can ``go across'' different maximal degree-2 paths of $R$ only $2^{\OO(k)}$ many times (because when going across different maximal degree-2 paths, either the path needs to intersect a $2^{\OO(k)}$-sized separator or a ``short'' maximal degree-2 path). Having any solution $\cal P$ satisfy this property helps in proving that when we push any solution $\cal P$ onto $R$, at least those parts of the paths in $\cal P$ that go across different maximal degree-2 paths are only $2^{\OO(k)}$ in number, and therefore they can be accommodated using only $2^{\OO(k)}$ parallel edges for each edge of $R$.

\bibliographystyle{splncs04.bst}
\bibliography{references,referencesISF,referencesRS,PaPaAlg,referencesPDP,referencesPDP1,referencesPDP2}

\end{document}